\documentclass[fleqn,usenatbib]{mnras}

\usepackage{newtxtext,newtxmath}
\usepackage[T1]{fontenc}

\DeclareRobustCommand{\VAN}[3]{#2}
\let\VANthebibliography\thebibliography
\def\thebibliography{\DeclareRobustCommand{\VAN}[3]{##3}\VANthebibliography}

\def\swift{{\em Swift}}
\def\grb{GRB\,210702A}

\usepackage{graphicx}	% Including figure files
\usepackage{amsmath}	% Advanced maths commands
\title[Rapid radio brightening of GRB 210702A]{Rapid radio brightening of GRB 210702A}

\author[G. E. Anderson et al.]{G. E. Anderson,$^{1}$\thanks{E-mail: gemma.anderson@curtin.edu.au}
T. D. Russell,$^{2}$
H. M. Fausey,$^{3}$
A. J. van der Horst,$^{3}$
P. J. Hancock,$^{4}$
A. Bahramian,$^{1}$
\newauthor
M. E. Bell,$^{5,6}$
J. C. A. Miller-Jones,$^{1}$
G. Rowell,$^{7}$
M. W. Sammons,$^{1}$
R. A. M. J. Wijers,$^{8}$
T. J. Galvin,$^{1,9}$
\newauthor
A. J. Goodwin,$^{1}$
R. Konno,$^{10}$
A. Rowlinson,$^{8,11}$
S. D. Ryder,$^{12,13}$
F. Sch\"ussler,$^{14}$
S. J. Wagner,$^{15}$
S. J. Zhu$^{10}$
\\
$^{1}$International Centre for Radio Astronomy Research, Curtin University, GPO Box U1987, Perth, WA 6845, Australia \\
$^{2}$INAF, Istituto di Astrofisica Spaziale e Fisica Cosmica, Via U. La Malfa 153, I-90146 Palermo, Italy\\
$^{3}$Department of Physics, George Washington University, 725 21st St NW, Washington, DC 20052, USA\\
$^{4}$Curtin Institute for Computation, Curtin University, GPO Box U1987, Perth, 6845, WA, Australia\\
$^{5}$University of Technology Sydney, 15 Broadway, Ultimo NSW 2007, Australia \\
$^{6}$The University of Newcastle, University Drive, Callaghan NSW 2308, Australia \\
$^{7}$School of Physical Sciences, The University of Adelaide, Adelaide
5005, Australia\\
$^{8}$Anton Pannekoek Institute for Astronomy, University of Amsterdam, Science Park 904, NL-1098 XH Amsterdam, The Netherlands \\ 
$^{9}$CSIRO Space \& Astronomy, PO Box 1130, Bentley WA 6102, Australia\\
$^{10}$Deutsches Elektronen-Synchrotron (DESY), D-15738 Zeuthen, Germany\\
$^{11}$ASTRON, the Netherlands Institute for Radio Astronomy, Postbus 2, NL-7990 AA Dwingeloo, The Netherlands\\
$^{12}$School of Mathematical and Physical Sciences, Macquarie University, NSW 2109, Australia \\
$^{13}$Astronomy, Astrophysics and Astrophotonics Research Centre, Macquarie University, Sydney, NSW 2109, Australia\\
$^{14}$IRFU, CEA, Universit\'e Paris-Saclay, F-91191 Gif-sur-Yvette, France\\
$^{15}$Landessternwarte, Universit\"at Heidelberg, K\"onigstuhl, D 69117 Heidelberg, Germany
}

\date{Accepted XXX. Received YYY; in original form ZZZ}

\pubyear{2023}

\begin{document}
\label{firstpage}
\pagerange{\pageref{firstpage}--\pageref{lastpage}}
\maketitle

% Abstract of the paper
\begin{abstract}

We observed the rapid radio brightening of GRB 210702A with the Australian Telescope Compact Array (ATCA) just 11\,h post-burst,
tracking early-time radio variability over a 5\,hr period on $\sim15$\,min timescales at 9.0, 16.7, and 21.2\,GHz. 
A broken power law fit to the 9.0\,GHz light curve showed that the 5\,h flare peaked at a flux density of $0.4 \pm 0.1$\,mJy at $\sim13$\,h post-burst. 
The observed temporal and spectral evolution is not expected in the standard internal-external shock model, where forward and reverse shock radio emission evolves on much longer timescales.
The early-time ($<1$\,d) optical and X-ray light curves from the \textit{Neil Gehrels Swift Observatory} demonstrated typical afterglow forward shock behaviour, allowing us to use blast wave physics to determine a likely homogeneous circumburst medium and an emitting electron population power-law index of $p=2.9 \pm 0.1$. 
We suggest that the early-time radio flare is likely due to weak interstellar scintillation (ISS), which boosted the radio afterglow emission above the ATCA sensitivity limit on minute timescales. 
Using relations for ISS in the weak regime, we were able to place an upper limit on the size of the blast wave of $\lesssim 6 \times 10^{16}$\,cm in the plane of the sky, which is consistent with the theoretical forward shock size prediction of $8 \times10^{16}$\,cm for GRB 210702A at $
\sim13$\,h post-burst.
This represents the earliest ISS size constraint on a GRB blast wave to date, demonstrating the importance of rapid ($<1$\,d) radio follow-up of GRBs using several-hour integrations to capture the early afterglow evolution and to track the scintillation over a broad frequency range.

\end{abstract}


\section{Introduction} \label{sec:intro}

Gamma-ray bursts (GRBs) bridge several aspects of multi-messenger astrophysics and even 50 years following their discovery they continue to push the boundaries of physics. 
A large fraction of the GRB population show a dichotomy based on the duration and spectral properties of the prompt gamma-ray emission \citep{kouveliotou93}: long and spectrally soft GRBs (LGRBs; $>2$\,s) likely resulting from the collapse of massive ($\gtrsim10M_\odot$) stars \citep{woosley93,kulkarni98,woosley06,woosley06heger} and short and spectrally hard GRBs (SGRBs; $<2$\,s), which are likely the merger of a binary neutron star or a neutron star and black hole binary \citep{lattimer76,eichler89,narayan92,mochkovitch93,abbott17grb}.  
There are also events that seem to bridge this gap as demonstrated by examples on either side of the $2$\,s divide \citep{zhang09}, including SGRBs with extended emission \citep[e.g. GRB 060614 with $\sim102$\,s duration;][]{gehrels06} and the recent discovery of the shortest GRB from a collapse \citep[GRB 200926A with $0.65$\,s duration;][]{ahumada21,zhang21}. 

The standard Fireball model for GRB emission includes the internal-external shock scenario \citep{sari97,piran99rev}, where the 
the gamma-ray emission is generated by processes internal to a relativistic jet, powered by rapid accretion onto a newly formed black hole or rapidly spinning neutron star
\citep{rees92,kobayashi97,sari99,kobayashi00sari}.
These jets interact with the circumburst medium (CBM) comprised of material ejected by the progenitor during its lifetime, sweeping up material that drive a blast wave that is observed as an afterglow from radio up to very-high-energy (TeV) gamma-rays \citep{paczynski93,meszaros97,wijers97,frail97,waxman97ApJ...489L..33W,abdalla19Nat,acciari19nat,hess21}. This afterglow consists of a forward shock that propagates outwards into the CBM, which then generates a short-lived reverse shock that propagates back into the shocked environment. 
Both shocks have been observed as two distinct synchrotron spectral components from multiple GRBs \citep[e.g.][]{laskar13grb130427a,perley14,vanderhorst14,laskar16grb160509a,laskar19grb181201a} and so the broadband afterglow is usually modelled under the reverse-forward external shock framework \citep{meszaros97}. 

The evolving synchrotron spectra from both the forward and reverse shock are approximated by power-law segments between three characteristic frequencies: the synchrotron self-absorption frequency $\nu_a$, the peak frequency $\nu_m$, and the cooling frequency $\nu_c$ \citep{sari98,wijers99,granot02}. By identifying these frequencies (of which 2 of the 3 can only be constrained through radio observations) and the peak flux $S_{\nu,max}$, it is possible to derive physical properties of the GRB outflow. 
The inclusion of radio observations in forward shock modelling reveals details on the total energy budget of the GRB, the outflow geometry, the CBM density structure, 
and the microphysics of the observed emission  
\citep[e.g.][]{vanderhorst08,vanderhorst14,alexander17}. Meanwhile, the reverse shock is sensitive to the jet composition (baryon content), the initial Lorentz factor and the magnetisation of the GRB jets \citep[e.g.][]{laskar19grb181201a,laskar19grb190114c}. 
While the forward shock dominates from optical frequencies and upwards from a few minutes post-burst, this component may take up to tens of days to peak in the radio band. 
However, depending on the CBM density profile, reverse shock radio emission can be detected peaking within 1-2\,d post-burst \citep[e.g.][]{kulkarni99,frail00,berger03,soderberg06,anderson14,laskar19grb190114c,lamb19,troja19}.  

While extremely rapid follow-up is possible in the gamma-ray, X-ray and optical regimes due to facilities such as the \textit{Neil Gehrels Swift Observatory} (\swift{}), 
the early-time radio properties of GRBs remain a poorly constrained regime. 
Rapid radio follow-up ($<1$\,day) is particularly crucial for catching the rapidly evolving reverse shock emission. 
Only a few radio telescopes are currently equipped with the ability to rapidly and automatically respond to external GRB triggers at GHz frequencies, including the Australia Telescope Compact Array \citep[ATCA;][]{anderson21atca} and the Arcminute Microkelvin Imager \citep[AMI;][]{staley13,anderson14,anderson18}. There are also only a few radio telescopes that are currently automatically triggering in the MHz regime \citep[e.g. the Murchison Widefield Array and the Low-Frequency Array;][]{kaplan15,rowlinson19grb,rowlinson21,anderson21mwa,tian22pasa,tian22mnras}, which are targeting more exotic models that predict prompt, coherent radio emission from a rapidly rotating and highly magnetised neutron star remnant that may be formed by both long and short GRBs \citep[see][for a review of some models]{rowlinson19mod}.

The typical distance to GRBs combined with the physical size of their jets makes them susceptible to interstellar scintillation (ISS) caused by inhomogeneities in the ionised interstellar medium, which distort the incoming wavefront causing temporal fluctuations in the radio light curve \citep{rickett90,narayan92iss,goodman97,walker98}. 
The type of scintillation depends on the observing frequency and the compactness of the source compared to the ISS characteristic angular scale.
As the GRB jet expands with time, the effect of ISS decreases until it is quenched. 

The detection of ISS in GRB radio light curves allows for a direct measure or upper limit on the size of the jetted outflow on the plane of the sky, which can be directly compared to source size predictions from afterglow modelling and is therefore a powerful test of the physical assumptions underlying the Fireball model \citep[e.g.][]{frail00scint,alexander19}. 
ISS has been observed for a few GRBs, which have resulted in size constraints of  $\lesssim 10^{17}$\,cm \citep[][]{frail97,frail00scint,chandra08,vanderhorst14,alexander19,rhodes22},
which are comparable and possibly more constraining than what is possible using very long baseline interferometry \citep[VLBI; e.g.][]{taylor04}.

There are also a few GRBs with radio afterglow evolution that deviates from the expected reverse-forward external shock framework for which ISS may not be the only explanation. 
A detailed radio analysis of GRB 160131A by \citet{marongiu22} detected several spectral peaks (the most consistent at 8\,GHz) in its radio spectra between 0.8 and 25\,days post-burst that could be due to ISS, a two-component jet \citep[e.g][]{vanderhorst14}, or a thermally emitting population of electrons \citep{eichler05}. Such a thermal population would be the fraction of electrons not accelerated into a power law distribution by the shock and may result in an early radio ``pre-brightening'' \citep{eichler05,giannios09,ressler17}. 
Another example includes GRB 130925A, which 2.2\,days post-burst showed a radio peak at 7\,GHz and spectral cut-off at $>10$\,GHz, which could be attributed to ISS or an unusual underlying electron population \citep[mono-energetic or an unusually steep power-law energy distribution,][]{horesh15}.
Such results demonstrate that there may be unknown physics occurring at early times, illustrating our need for rapid and comprehensive radio follow-up of GRBs. 

The early onset of radio afterglows, the observed $\sim1-2$\,day evolution of the radio reverse shock, the detection of unusual early-time radio spectral features, and the potential for using ISS for source size estimates are all strong arguments for supporting rapid and even automated/triggered radio follow-up of GRBs with several hours of integration. 
GRB radio afterglows are shown to become detectable by current instrumentation between $4-16$\,hr post-burst \citep[e.g.][]{anderson18} so a $\leq12$\,hr integration at early times could allow us to track the rise in the radio afterglow components and ensure periods of quasi-simultaneous X-ray, UV and optical follow-up provided by standard GRB monitoring. 
As a result, our team has conceived a rapid-response observing program with ATCA to perform triggered follow-up of GRBs detected by \swift{} to probe the early-time physics of these extremely energetic transients \citep[see][for a description of the triggering strategies and software]{anderson21atca}. 

For this paper, we present the rapid radio brightening observed from GRB 210702A by ATCA, which is not typical afterglow behaviour and likely stems from ISS. 
In Section 2, we describe the ATCA and \swift{} datasets and reduction, which is followed by an analysis of the radio variability and spectral properties in Section 3. In Section 4, we discuss the radio, optical and X-ray emission in the context of the afterglow (internal-external shock) model and ISS. We then compare our ISS-derived source size estimates to those predicted for the forward shock. We summarise our conclusions in Section 5 to promote how similar rapid early-time observations could be exploited in the future to constrain source size and thus test the underlying assumptions of GRB afterglow modelling.

\section{Observations and Processing} \label{sec:obs}

The \swift{} Burst Alert Telescope (BAT) detected GRB 210702A at 19:07:13 UT (trigger number 1058804) on 2021-07-02 \citep{lien21}, and it was localised at X-ray and optical wavelengths with the \swift{} X-ray Telescope \citep[XRT;][]{burrows05} and the Ultra-violet Optical Telescope \citep[UVOT;][]{roming05}, with the best final \swift-UVOT position at $\alpha \mathrm{(J2000.0)} =  11^{\mathrm{h}}14^{\mathrm{m}}18\overset{\mathrm{s}}{.}70$ and $\delta (\mathrm{J2000.0}) = -36^{\circ}44'50\overset{''}{.}0~$ with a 90\% confidence error circle of $0\overset{''}{.}46$ \citep{kuin21}. Spectra of the optical afterglow obtained by the European Southern Observatory's Very Large Telescope X-Shooter spectrograph \citep{vernet11} found a likely redshift of $z=1.160$ \citep{xu21}, which is broadly consistent with that obtained rapidly by \swift{}-UVOT \citep{kuin21z}. The bright optical and gamma-ray counterpart was also monitored by several other facilities. 

Several radio detections were reported within the first few days post-burst, including $\sim0.1$\,mJy at 97.5\,GHz with the Atacama Large Millimeter/Submillimeter Array \citep[ALMA;][]{wootten09} at 26.0\,h post-burst \citep{laskar21alma_early} and with ATCA at 16.7, 21.1 and 34\,GHz at 3.5\,days post-burst \citep[reporting $\sim0.8$\,mJy at 34\,GHz;][]{laskar21atca_early}. The radio afterglow significantly faded over the first $\sim10$\,days post-burst only to be shown to  have rebrightened contemporaneously at $\sim18$\,days post-burst with ATCA at 5.5, 9.0, 16.7, 21.2 and 34\,GHz \citep[][]{laskar21atca_late} and with ALMA at 97.5\,GHz \citep[][]{laska21alma_late}. The MeerKAT Radio Telescope \citep{jonas16} also detected a brightening radio counterpart at 21.0\,days post-burst with a flux density of $1.23 \pm 0.12$\,mJy at 1.284\,GHz \citep[][]{laskar21meerkat}.  

The late-time radio rebrightening is an unusual light curve feature for GRBs seen from few events. One such example includes GRB 050416a, which produced a bright radio flare at $\sim40$\,days post-burst.
Possible explanations include a very sudden increase in CBM density or energy injection from either a refreshed shock (caused by a slower ejecta shell catching up with the afterglow shock) or an off-axis ejecta \citep{soderberg07}. 
However, the focus of this paper is on another unusual radio light curve feature at early times during which ATCA detected minute timescale variability from GRB 210702A at 5.5, 9.0, 16.7 and 21.1 GHz between 9-14 hrs post-burst.

\subsection{ATCA}\label{sec:obs_atca}

The ATCA observations of GRB 210702A began at 2021-07-03 00:31:30 UT 
when the target rose above the horizon just 5.4 hrs post-burst. 
The ATCA radio observations were conducted manually as there was enough time to respond between the \swift{} trigger and the target rising, which occurred during the observer's daytime. 
However, this observation demonstrates the potential of rapid-response systems for ensuring we are on-target to detect early-time radio emission from a large number of events without relying on human response. 
The observations were centred on the initial \swift{}-UVOT position of $\alpha \mathrm{(J2000.0)} =  11^{\mathrm{h}}14^{\mathrm{m}}18\overset{\mathrm{s}}{.}83$ and $\delta (\mathrm{J2000.0}) = -36^{\circ}44'48\overset{''}{.}8~$ with a 90\% confidence error circle of $0\overset{''}{.}61$ \citep{lien21} and 
ATCA observed until 11:29:20 UT, totalling 11\,hr on source. 
Observations were conducted in the 6B configuration (an extended 6\,km configuration)\footnote{\url{ https://www.narrabri.atnf.csiro.au/operations/array_configurations/configurations.html}} and alternated between the two dual ATCA receivers with central frequencies at 5.5/9.0 GHz and 16.7/21.2 GHz, which all have 2 GHz bandwidths. During the first 3.7\,h of observation, antenna 2 was unavailable, resulting in a loss of 5 of the usual 15 baselines, but was back online for the remainder of the 11 hours. 

The ATCA data were reduced using the radio reduction software Common Astronomy Software Applications package \citep[CASA;][]{casa22pp}, version 5.1.2, using standard techniques. Flux and bandpass calibration were performed using PKS 1934--638, with phase calibration using interleaved observations of PKS 1144-379. 

The radio afterglow was clearly detected in the 11\,hr integration at 9.0, 16.7 and 21.2\,GHz, with the best position of $\alpha \mathrm{(J2000.0)} =  11^{\mathrm{h}}14^{\mathrm{m}}18\overset{\mathrm{s}}{.}81$ and $\delta (\mathrm{J2000.0}) = -36^{\circ}44'49\overset{''}{.}3~$ with a 90\% confidence error circle of $0\overset{''}{.}3$, which is consistent within $3\sigma$ of the best \swift-UVOT position \citep{kuin21}. 
There was no detection at 5.5\,GHz in the full 11\,hr image with a $3\sigma$ upper limit of $27\mu$Jy. 
Given that there is likely radio afterglow variability at early times sometime between $4-16$\,h post-burst \citep{anderson18}, we further investigated for evidence of intra-observational variability (see Section~\ref{sec:var}). 

\subsection{Swift observations}\label{sec:obs_swift}

The \swift-XRT and -UVOT performed follow-up observations of GRB 210702A. \swift-XRT observations began at 85\,s post-burst and were obtained up to 10\,d post-burst. These data were automatically processed and accessed via the online catalogue of XRT results \citep{evans07,evans09}, which includes light curve and spectral modelling of GRB 210702A that are summarised in the following. 
Throughout this paper we assume the flux density is represented by the power law $S_{\nu} \propto t^{\alpha} \nu^{\beta}$ for all wavelength bands.

The \swift-XRT light curve is best described by a broken power law fit with temporal indices  of $\alpha_{X,1}=-0.979_{-0.017}^{+0.014}$ and $\alpha_{X,2}=-1.452_{-0.029}^{+0.024}$ on either side of a break at $4416^{+757}_{-487}$\,s. No flares were present in the X-ray light curve. 
At the time of our radio detection, the X-ray light curve was in its steeper declining phase. We therefore assume the X-ray spectrum was best described by an absorbed power law fit that was performed using $\sim3$\,h of data centred at 4.5\,h post-burst, reporting a Galactic and intrinsic absorption column of $N_{H}=1.19 \times 10^{21}$\,cm$^{-2}$ and $2.3^{+1.0}_{-0.9} \times 10^{21}$\,cm$^{-2}$, respectively, and a photon index of $\Gamma=1.95 \pm 0.06$ \citep{evans09}. The X-ray spectral index is related to the photon index by $\beta_X = 1 - \Gamma$ so at the time of our radio observations $\beta_X=-0.95 \pm 0.06$. 

The \swift-UVOT monitored GRB 210702A between 200\,s and 12\,days post-burst in the $UW2$, $UW1$, $U$, $B$, and $V$ filters. 
These data were obtained from The Neil Gehrels \emph{Swift} Observatory Data Archive\footnote{\url{https://www.swift.ac.uk/swift_portal/}}. We processed the data using \textsc{HEASoft} \citep[version 6.29,][]{heasoft14}, and extracted high-level UVOT products through \texttt{uvotproduct}, performing photometry using a 5$''$ source extraction region, and a background annulus region (centered on the source) with inner and outer radii of 12.5$''$ and 25$''$ (with 6$''$ exclusion regions for any sources detected within the background region). We also checked for Small Scale Sensitivity issues following standard procedures \footnote{\url{https://swift.gsfc.nasa.gov/analysis/uvot_digest/sss_check.html}} and disregard any affected data.
A power law fit to the UVOT light curves in each filter band resulted in a temporal index of $\alpha_{O} \sim -1.4$, which is similar to the XRT light curve decline following the X-ray break. 
 
\section{Variability and spectral analysis}\label{sec:atca_lc_spec}

\subsection{Intra-observational radio variability}\label{sec:var}

Having identified the radio counterpart, we searched for intra-observational variability in the hope of tracking the rise of the afterglow at each observing frequency.
Due to the elongated beam produced by the ATCA East-West array configuration on short timescales, we fitted for a point source in the visibility plane using \textsc{uvmultifit} \citep{2014A&A...563A.136M}. 
To best represent the observed flux density variations, we explored different variability timescales at each frequency band, with intervals between 5 minutes and 1-hour. 
For all detections, the fitting position was left free and source flux densities were measured by fitting for a point source in the uv-plane. 
We chose to leave the fitting position free as for such short integrations, the ATCA beam becomes very elongated, particularly at higher frequencies. However, the scatter from the true position was always within the beam.
For the case when the \grb\ was not detected (a signal-to-noise or SNR$<$3), we fixed the fitting position to the best known source position from UVOT (see Section~\ref{sec:obs}).
This resulted in a force-fitted flux density and error measurement rather than just $3\sigma$ upper limit, which has the advantage of accounting for the presence of nearby sources and noise variations across the image.
The errors from \textsc{uvmultifit} are determined
from the Jacobian matrix and then scaled so that the reduced Chi-square equals unity. These errors were found to be similar to the rms
of an image over the same integration time. As such, our $3\sigma$ detection
threshold was treated as 3 times the \textsc{uvmultifit} error.

\begin{figure}
    \centering
    \includegraphics[width=0.48\textwidth]{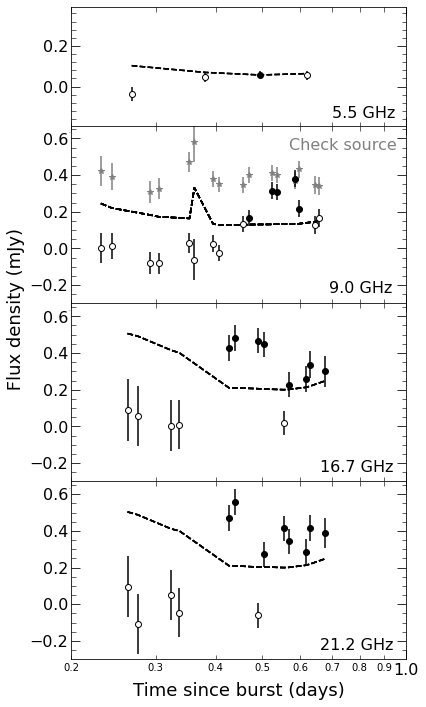}
    \caption{Radio light curves of GRB 210702A at 5.5 (60\,min bins), 9.0 (15\,min), 16.7 (12.5\,min) and 21.2\,GHz (12.5\,min). Filled data points are considered detections (SNR$\geq3$) whereas the hollow circles are force-fitted flux density values (SNR$<3$). The dashed line represents the detection threshold (marked as an SNR of 3).
    At 9.0\,GHz we also plot the flux density of a check source in the field to demonstrate that the observed transient nature of what we have identified as the radio afterglow to GRB 210702A was not an instrumental artefact.
    }
    \label{fig:atca_lc}
\end{figure}

\begin{figure}
    \centering
    \includegraphics[width=0.48\textwidth]{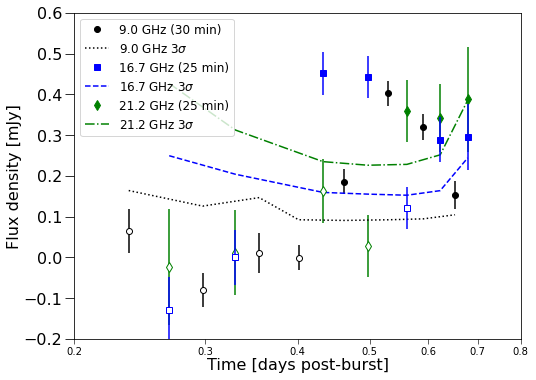}
    \caption{Lower cadence radio light curves of GRB 210702A at 9.0 (30\,min), 16.7 (25\,min) and 21.2\,GHz (25\,min) to demonstrate the late-time detections are significant compared to the less sensitive early-time data and therefore indicative of transient behaviour. Filled data points are considered detections (SNR$\geq3$) whereas the hollow circles are force-fitted flux density values (SNR$<3$). The dashed, dash-dotted and dotted lines represents the detection threshold at each of the three plotted frequencies (marked as an SNR of 3).}
    \label{fig:atca_lcad}
\end{figure}

The light curves are shown in Figure~\ref{fig:atca_lc} where the scans at each frequency were binned on timescales of 60, 15, 12.5 and 12.5 minutes at 5.5, 9.0, 16.7 and 21.2\,GHz, respectively. 
The plotted flux density measurements at each frequency are listed in Table~\ref{tab:radio_flux}. 
The $3\sigma$ detection thresholds are shown as a dashed line in each panel. 
The noise is noticeably higher at early times due to the absence of antenna 2. We therefore double the bin size at 9.0, 16.7 and 21.2\,GHz (see Figure~\ref{fig:atca_lcad}) to demonstrate that the late-time detections are significant compared to the less sensitive early-time data and therefore indicative of a transient nature. 

To quantify the variable nature of the radio counterpart we have identified, we performed a $\chi^{2}$ probability test that the flux density light curve variations were consistent with a steady source following the technique outlined in section 4 of \citet{bell15}, who define a source as variable if $P<0.001$. For the 9, 16.7 and 21.2\,GHz light curves in Figure~\ref{fig:atca_lc}, we calculated probabilities of $P=2\times10^{-19}$, $3\times10^{-7}$, and $1\times10^{-10}$, respectively, confirming they are variable according to the above criterion.

At 9.0\,GHz, all other sources in the field but one `check' source were uv-subtracted using the \textsc{casa} task \texttt{uvsub}. Here, the check source was left in to test that the observed variability was not an instrumental artefact \citep[e.g.][]{frail00scint}. No significant variability was detected in the check source, with the $\chi^{2}$ returning a probability of $P=0.6$ of being a steady source (see Figure~\ref{fig:atca_lc}). 
We note the sudden increase in the noise in the 9.0\,GHz light curve at 8.7\,h (0.36\,d) post-burst, which is due to that particular scan being cut short to 4\,min (from 15\,min) when the observations were paused to reintegrate antenna 2 back into the array. This caused the flux density measurement of the check source to increase, however, the corresponding increase in error bars shows it was consistent within $2\sigma$ of the other measurements. 

Due to the lack of multiple detections of \grb, we did not do the same check test at 5.5\,GHz. In our 16.7 and 21.2\,GHz data, there was no check source sufficiently bright to follow the same steps. As such, to ensure that the variability at these frequencies was intrinsic to the source, we re-analysed these higher frequency ATCA data treating every second phase calibrator scan as the `target', and every other phase calibrator scan as the calibrator. Doing this did not show any significant changes in the `target' phase calibrator scans. Hence, we conclude that the observed flux variations of \grb\ are intrinsic to the source.

The most prominent light curve feature is the flare at 9.0\,GHz, which becomes detectable at 11\,h post-burst, peaks, and then fades away to nearly below detectability towards the end of the observation at 15.6\,h post-burst.
Variability is also observed at 5.5, 16.7 and 21.2\,GHz but with these frequencies showing less structured behaviour (see Figure~\ref{fig:atca_lc}).

To better characterise the flaring behaviour at 9.0\,GHz, we fit a broken power-law to the light curve using the MCMC Python package {\sc pymc3} \citep{salvatier16} according to the following model:
\begin{equation}\label{eq:bkpl}
    S = \begin{cases} A (t/T)^{\alpha_{1}},&  t \leq t_{b}  \\
    A (t_{b}/T)^{(\alpha_{1}-\alpha_{2})} (t/T)^{\alpha_{2}},& t > t_{b}
    \end{cases}
\end{equation}
where $T$ is the mean time of the 9.0\,GHz measurements. 
We assume uniform priors when fitting for the break time $t_{b}$, amplitude $A$, and the temporal indices $\alpha_{1}$ and $\alpha_{2}$, constraining the latter to be positive and negative, respectively. 
Rather than excluding the non-detections from our modelling, we included all the 9\,GHz force-fitted flux density measurements for which the SNR$<3$ in the broken power law fit as we could assign a likelihood to a predicted model flux for a set of model parameters (something that is not possible with an upper limit). Examples of where radio force-fitted flux densities are reported and used in afterglow modelling include \citet{galama98,kulkarni99,vanderhorst11,vanderhorst15}.
The mean broken power law fit and a sample of traces at 9.0\,GHz are shown in the top panel of Figure~\ref{fig:bpl_spec_index} (see the corresponding corner plots in Figure~\ref{fig:corn_9ghz}). 
The resulting temporal indices show a steep rise of $\alpha_{1}=+7.3^{+1.7}_{-1.3}$ until $13.4 \pm 0.2$\,h ($0.56 \pm 0.01$\,days) when the emission rapidly decays with $\alpha_{2}=-8.4 \pm 3.7$ ($1\sigma$ errors). 
Such temporal indices are extremely steep and not characteristic of observed GRB radio afterglows. We explore this further in Section~\ref{sec:afterglow}. 
The radio emission appeared to become detectable progressively earlier with increasing frequency as demonstrated by the 16.7 and 21.2\,GHz light curves in Figure~\ref{fig:atca_lc}. At both frequencies they also briefly dropped below detectability at 13.4\,h (0.56\,d) and 11.8\,h (0.49\,d) post-burst, respectively, before brightening again, which could be the rise of a second flare. The more erratic variability meant that broken power law fits at these two frequencies did not converge.

As a further test of the nature of the radio emission, we calculated the brightness temperature ($T_b$) that such rapid variability would imply if it were intrinsic to the source. Using equation 1 in \citet{anderson14}, we took the first and second detections at 9.0 GHz (where the non-fixed position fit returned a SNR$>3$, see filled data points in  Figure~\ref{fig:bpl_spec_index}) and calculated $T_{b}=2.3 \times 10^{17}$\,K and a Lorentz factor of $\Gamma=65$. Meanwhile, the last two detections give $T_{b}=6.2 \times 10^{18}$\,K and a Lorentz factor of $\Gamma=183$. 
Such high values of $T_{b}$ and $\Gamma$ are not typical of those derived from radio detections of GRBs within $\sim1$\,day post-burst \citep[$T_{b}\sim10^{15}-10^{16}$\,K and $\Gamma \sim 10-20$;][]{anderson18} but have been observed from GRB 161219B and attributed to diffractive interstellar scintillation \citep{alexander19}.

\subsection{Radio spectral analysis}\label{sec:spec}

\begin{figure}
    \centering
    \includegraphics[width=0.48\textwidth]{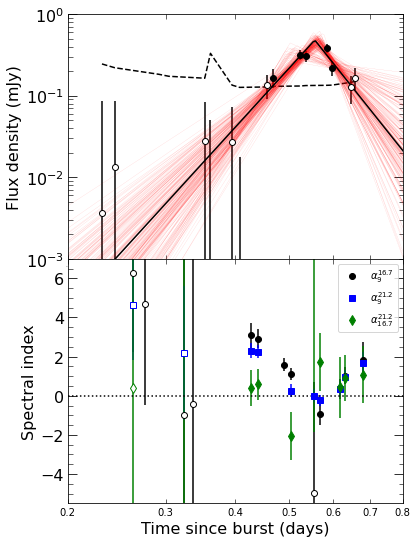}
    \caption{Top: Broken power law fit to the 9.0\,GHz light curve. Data points are the same as for Figure~\ref{fig:atca_lc}. The red lines show a random selection of 200 out of 5000 trace samples from the MCMC model fitting. See the corresponding corner plots in Figure~\ref{fig:corn_9ghz}. 
    Bottom: Radio spectral index temporal evolution between frequency pairs 9.0/16.7\,GHz, 9.0/21.2\,GHz, and 16.7/21.2\,GHz. The broken powerlaw model traces of the 9.0\,GHz light curve were used to interpolate the 9.0\,GHz flux to the time of the 16.7 and 21.2\,GHz observations. Filled data points are spectral indices calculated using the detections (SNR$\geq3$) at 16.7 and 21.2\,GHz whereas open circles show the spectral index calculated using the force-fitted flux density values  (SNR$<3$).  }
    \label{fig:bpl_spec_index}
\end{figure}

In order to explore the spectral behaviour of the radio emission during the observed flare, 
we used the broken power law fit of the 9.0\,GHz light curve derived in Section~\ref{sec:var} to interpolate the 9.0\,GHz flux density to the time of the 16.7 and 21.2\,GHz measurements, using the traces (red lines in Figure~\ref{fig:bpl_spec_index}) to estimate the $1\sigma$ error.
We also calculated the spectral index between the 16.7 and 21.2\,GHz measurements as they were simultaneously obtained with the 15\,mm receiver.
The evolution of these three pairs of spectral indices are shown in the bottom panel of Figure~\ref{fig:bpl_spec_index}. 

During the flare observed at 9.0\,GHz, the spectral index of the 9.0\,GHz frequency pairs evolved from $\beta \sim +3$ to $\sim0$ and back up to $\sim+2$, while the evolution between 16.7 and 21.2\,GHz was less extreme and remained closer to $\beta \sim 0$. 
While we expect the spectral index to be $\beta \sim+2$ at early times from both the forward and reverse shock emission in the radio band before the observing frequency drops below the synchrotron self-absorption frequency, we also expect it to remain steady on day(s) timescales \citep[see][for a summary of analytical synchrotron external shock GRB models]{gao13}. 
This prediction is very different to the extreme spectral variability we observe over a 5 hour period from GRB 210702A (see discussion in Section~\ref{sec:afterglow}). 

\subsection{Optical to X-ray spectral analysis}\label{sec:swift_spec}

We used the \swift-UVOT and -XRT data to determine the optical-to-X-ray spectral energy distribution (SED) of GRB 210702A and accounted for extinction contributions from the Milky Way and host galaxy, as well as intergalactic attenuation.
We first performed a linear interpolation of the log-log light curves of the \swift-UVOT $U$, $V$, $B$, $UW1$, and $UW2$ filters using \texttt{scipy.interpolate.interp1D} \citep{scipy}, to estimate their flux at the common time of $T0 + 14370$\,s ($\sim4$\,h post-burst; the logarithmic average of the closest measurement times of each of the filters).
The fluxes for each filter are corrected for Milky Way extinction using $A_{\lambda}$ values from \citet{kataoka2008}. 
Similarly, we also interpolated the \swift-XRT data to estimate the flux in the 0.3-10 keV band at $T0 + 14370$\,s, which was also corrected for Milky Way and host galaxy extinction using the ratio of the unabsorbed and absorbed fluxes listed in the \swift-XRT GRB light curve repository \citep{evans07, evans09}. 
We used the measured \swift-XRT spectral index $\beta_{X} = -0.95 \pm 0.06$ (see Section~\ref{sec:obs_swift}) to determine the spectral flux density at the logarithmic average energy of the band, 1.73 keV (corresponding to $7.2$\AA). 
For all the interpolated data points, we took the largest percentage error on either of the surrounding points and applied that to the flux to ensure we were including a sufficient margin of error.

\begin{figure}[h!]
    \centering
    \includegraphics[width = 0.48\textwidth]{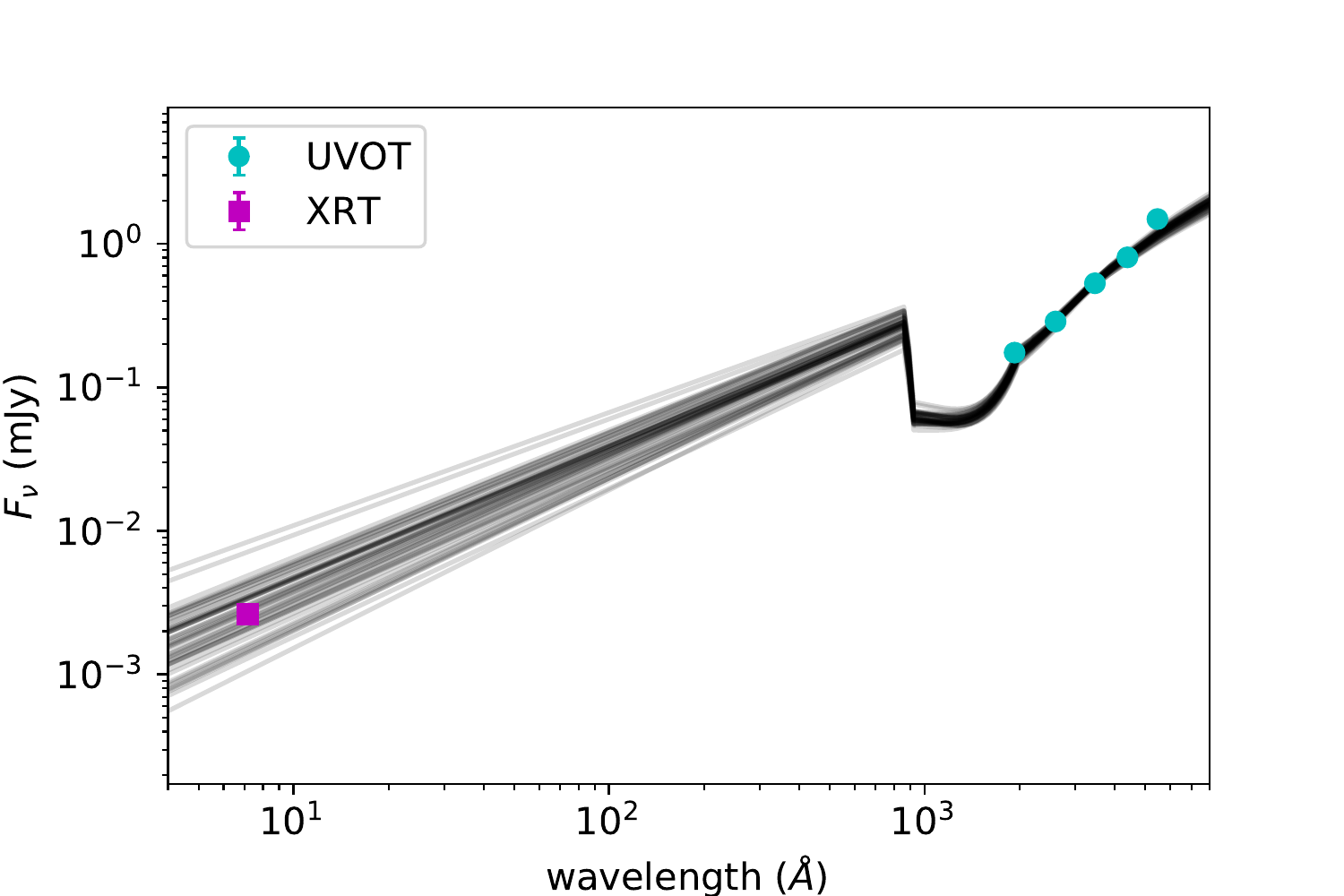}
    \caption{
    The optical-to-X-ray SED spectral fits for $F_\nu$ (mJy) as a function of wavelength (\AA) for the \swift-XRT and -UVOT data at $T0 + 14370$ s. 
    This log-log plot shows the final position of each of the 50 walkers for a Gaussian spectral index prior as described in Section~\ref{sec:swift_spec}. See the corresponding corner plots in Figure~\ref{fig:corn_xosed}.}
    \label{fig:end_walkers}
\end{figure}

Given that the light curve behavior demonstrates that the optical, UV and X-ray bands were in the same afterglow spectral regime (similar temporal indices, see Section~\ref{sec:obs_swift}), we assumed a single power-law spectrum for the SED modelling.
The fit included intergalactic attenuation, and a mix of photoelectric absorption and resonant scattering by hydrogen gas using a model by \cite{meiksin2006}, which provides updates and improvements to a previous model created by \citet{madau1995, madau1996}.
We also incorporated host galaxy extinction by using the Small Magellanic Cloud (SMC) model from \citet{pei1992} for the \swift-UVOT bands only as the X-ray data point already included a correction for host galaxy extinction.
Note that there is not enough of a spectral lever arm across the UVOT filters to be able to place constraints on an optical spectral index as well as the extinction. We therefore did not derive an optical spectral index.
The final optical and X-ray spectral values were fit using a Markov-Chain Monte-Carlo (MCMC) method using the \texttt{emcee} package \citep{foreman-mackey13emcee} to determine the best fit parameters for the amplitude, spectral index, and host galaxy extinction. 
The final positions of the walkers are plotted in Figure~\ref{fig:end_walkers}.
The fitting method includes a Gaussian prior for the X-ray-to-optical spectral index based on the measured X-ray spectral index value $\beta_{X} = -0.95 \pm 0.06$ (see Section~\ref{sec:obs_swift}).
This methodology provides posterior distributions for the parameters with peak posterior values and $1\sigma$ uncertainties of  %$A_{\lambda} = 3.74 \pm ^{0.45}_{0.40}$, 
$\beta_{X,O} = -0.96 \pm 0.06$ for the X-ray to optical spectral index and $E_{B-V} = 0.069 \pm 0.009$ for the host galaxy extinction (see the corresponding corner plots in Figure~\ref{fig:corn_xosed}).
The posteriors for both parameters exhibit a Gaussian distribution.
Given that the light curves had the same slopes in all the optical and X-ray bands from 4 to 12\,h post-burst, the SED did not change significantly over the time frame of the observed variable radio emission described in Section~\ref{sec:var}. 

\section{Discussion}

In the following we discuss the nature of the early-time radio flare. We first inspect our radio detections in the context of the internal-external shock scenario to explore the possibility of the observed variability being intrinsic to the GRB afterglow. We then expand our investigation to include extrinsic phenomena and discuss the implications of such early-time radio detections of GRBs in the future. 

\subsection{GRB afterglow}\label{sec:afterglow}

We first investigate whether the observed radio flare could be intrinsic to the GRB afterglow. 
At $\sim11$\,h post-burst, we expect the forward shock characteristic frequencies to be ordered as $\nu_a<\nu_m<\nu_c$, 
and that the optical and X-ray emission are dominated by this component. 
Given that the X-ray and optical temporal and spectral indices are similar, they are likely within the same power-law regime of the synchrotron spectrum. 
At $\sim1$\,d post-burst, we expect either  $\nu_{a,m}<\nu_{X,O}<\nu_c$ or $\nu_{a,m}<\nu_c<\nu_{X,O}$ so using the temporal ($\alpha$) and spectral ($\beta$) indices in both the X-ray and optical bands (see Section~\ref{sec:obs_swift} and \ref{sec:swift_spec}) we can use the closure relations derived by \citet{starling08} to calculate the power law distribution of the emitting electron population ($p$) and the index of the density profile of the CBM ($n \propto r^{-s}$, where $r$ is the distance from the progenitor).

At such early times, we are likely observing the GRB afterglow emission before any jet break, and thus the break in the X-ray light curve at 4416\,s is due to the cessation of an energy injection process \citep{zhang06}.
If we first investigate the scenario where $\nu_{a,m}<\nu_c<\nu_{X,O}$, then the closure relations using the spectral and temporal indices give different values for $p$; $p=-2\beta=1.90 \pm 0.12$ and $p=(-4\alpha+2)/3=2.60 \pm 0.04$, respectively. 
While there have been some cases where $p<2$, the mismatch between the values of $p$ derived from the spectral and temporal indices  suggest the X-ray and optical bands were in a different spectral regime.
When examining the case where $\nu_{a,m}<\nu_{X,O}<\nu_c$
we derive $p=-2\beta+1=2.90 \pm 0.12$. 
In this regime, the relationship between $\alpha$ and $p$ is dependent on the CBM density profile.
If we assume the CBM is homogeneous ($s=0$) 
then we derive $p=(-4\alpha+3)/3 = 2.94 \pm 0.04$, and for a wind medium ($s=2$) we derive $p=(-4\alpha+1)/3 = 2.27 \pm 0.04$. 
Based on these considerations it is likely that the X-ray and optical bands at the time of our radio detection were below the cooling break, the GRB CBM is likely homogeneous ($s=0$), and that $p=2.9 \pm 0.1$.  
While $p=2.9$ is on the higher end of the known distribution \citep[$2\lesssim p \lesssim 3$;][]{starling08,ryan15}, it is in no way unprecedented as \citet{starling08} derived a similar value for GRB 980519 ($p=2.96^{+0.06}_{-0.08}$) that was best modelled by a homogeneous medium ($s=0.23^{+1.22}_{-3.05}$) at 22.3\,h (0.93\,d) post-burst when $\nu_X<\nu_c$.

We also investigate whether the break in the X-ray light curve could be due to a jet break. Post-jet break, the value for $p$ is independent of the spectral regime and CBM density where $\alpha=p$ \citep{sari99jetS}. 
However, the shallow post-break temporal slope is too shallow: it would imply $p=1.4$, which is physically unlikely \citep[we would expect $p\sim2$;][]{sari99jetS} and inconsistent with the values of $p$ derived via closure relations using $\beta$ \citep{starling08}. 
Instead, this early break likely signifies the end of a plateau or energy injection phase, and the transition to a `normal' afterglow, which is often seen in early \swift{} X-ray light curves \citep{zhang06}.

At such early times, we may expect contributions from both the forward and reverse shock at radio wavelengths. For example, one of the closest known events GRB 030329 was detected at 14\,h ($\sim0.6$\,d) post-burst at 8.46\,GHz \citep{berger03nat} where late-time radio modelling demonstrated the radio afterglow was dominated by the forward shock \citep{vanderhorst08}. Meanwhile, other early-time ($<1$\,day post-burst) radio detections of GRBs are of the rapidly evolving reverse shock emission \citep[e.g.][]{kulkarni98,frail00,soderberg07,anderson14,vanderhorst14,laskar16grb160509a,laskar19grb181201a,laskar19grb190114c,lamb19,troja19}. We therefore consider both scenarios for the observed radio emission from GRB 210702A by comparing our 9.0\,GHz light curve and spectral index evolution to the steepest temporal and spectral index predictions for both the forward and reverse shock derived from analytical synchrotron external shock models of GRBs \citep{gao13}. 
Limiting this comparison to the $\nu_a<\nu_m<\nu_c$ spectral regime for a homogeneous medium and $p>2$ (based on our X-ray and optical results), we might expect 
the steepest rising power law segments for a thin shell (Newtonian) forward shock model of $S_{\nu}\propto t^{2}\nu^{2}$ and $S_{\nu}\propto t^{3}\nu^{2}$ for $\nu<\nu_a$ and $\nu_a < \nu <\nu_m$, respectively. 
The reverse shock (assuming the same environment and properties as above) rises even more steeply, with a flux density evolution of $S_{\nu}\propto t^{5}\nu^{2}$ for $\nu<\nu_a$, which may eventually change to $S_{\nu}\propto t^{1/3}\nu^{-1/2}$ as $\nu_a$ drops below the radio band.
 
The broken power law fit to the 9.0\,GHz light curve and radio spectral analysis shown in Figure~\ref{fig:bpl_spec_index} demonstrates the radio evolution is initially described by a very steep brightening and initial spectral index of $S_{\nu}\propto t^{7}\nu^{3}$. 
This temporal index agrees within $2\sigma$ and $4\sigma$ of the thin shell reverse and forward shock solution, respectively, for a homogeneous medium when $\nu<\nu_a$. 
However, the temporal index quickly swaps to $S_{\nu}\propto t^{-8}$ around $\sim13-14$\,h ($\sim0.6$\,d) post-burst, evolving to a spectral index of 0 and then back up to $\alpha\sim2$ at $\sim17$\,h ($\sim0.7$\,d). 
The errors on the second temporal index are less constraining so it agrees within $3\sigma$ of the $\nu_a < \nu <\nu_m$ forward and reverse shock temporal indices listed above. 
Regardless, the GRB afterglow synthetic light curves do not show the afterglow to be fading in this regime for nearly all the power law segments for any reasonable CSM profile or jet model assumptions \citep{gao13}.
 
Overall, while the optical and X-ray light observations show very typical forward shock behaviour at the time of our radio observations, the observed radio flare is not typical radio afterglow behaviour. This is also supported by the nonphysically high brightness temperature derived from the 9.0\,GHz light curve in Section~\ref{sec:var}. 
We therefore conclude that the radio variability is unlikely to be intrinsic to the reverse or forward shock emission.

\subsection{Scintillation}

We next explore whether the observed early-time radio variability from GRB 210702A could be due to ISS. 
For a given line of sight, inhomogeneities in the Milky Way are described by a scattering measure ($SM$) calculated from a free electron density model derived from pulsars \citep{taylor93,cordes02} or H-alpha maps \citep{hancock19pp}. The inhomogeneties are then modelled as a thin phase screen with a distance $d_{scr}$ in kpc, which in conjunction with the $SM$ is used to calculate a transition frequency $\nu_0 = 10.4 (SM_{-3.5})^{6/17}d_{scr}^{5/17}$\,GHz between the weak and strong scattering regimes \citep[where $SM_{-3.5}=SM/(10^{-3.5} \mathrm{kpc\,m}^{-20/3})$;][]{goodman97,walker98}. 
The size of the scattering disk responsible for the modulations can then be expressed in terms of the Fresnel scale  
$\theta_F=\sqrt{(\lambda d_{src})/(2 \pi)}$, 
$\nu_0$, and the observing frequency $\nu$. 
For $\nu>\nu_0$, the observations are in the weak scattering regime where the light curve modulations are broadband and caused by small phase changes on the Fresnel scale due to density changes in the ISM. As the weak regime is dominated by the Fresnel scale, the size of the scattering disk ($\theta_{scatt}$) is the size of the Fresnel scale \citep{narayan92iss}. 
For $\nu<\nu_0$, the observations are in the strong scattering regime where the wave front can experience two kinds of variability on different timescales. These include refractive scintillation, which is caused by broadband focusing and defocusing of the wave front by large-scale inhomogeneities in the scattering screen  \citep{sieber82,rickett84}. 
The other is diffractive scintillation, which is a narrow-band effect due to the interference between rays that are diffracted by small-scale inhomogeneities and thus operates over shorter timescales then refractive scintillation \citep{scheuer68}.
For a scintillation review see \citet{rickett90,narayan92iss}.

At 9.0, 16.7 and 21.2\,GHz, we use all flux density measurements with a SNR$\gtrsim3$ to measure the modulation index ($m = \sigma/\mu$, where $\sigma$ and $\mu$ are the standard deviation and mean) and roughly estimate the scintillation timescale ($t_{scint}$). Note there were not enough detections to constrain meaningful values at 5.5\,GHz. 
Given the very clear rise, peak and decline in the emission at 9.0\,GHz, it is possible we observed a single predominant scintillation cycle where we estimate the timescale to be $\sim5$\,h (the time between the first and last flux density measurement with an SNR$\gtrsim3$, see Figure~\ref{fig:bpl_spec_index}).
It is more difficult to identify a dominant timescale at both 16.7 and 21.2\,GHz due to the large error bars on the 12.5\,min measurements.
However, the non-detections we see at
13.4\,h (0.56\,d) and 11.8\,h (0.49\,d) post-burst at 16.7 and 21.2\,GHz, respectively, do imply the presence of short term variability. 
Given the $3\sigma$ detection threshold (see the dashed line in Figure~\ref{fig:atca_lc}) indicates the observations were sensitive enough to detect a source as bright as the significant neighbouring (before and after) detections.
We therefore estimate the scintillation timescale at 16.7 and 21.2\,GHz to be between the first detection at $9.6$\,h (0.4\,d) post-burst and the non-detection seen in both light curves, which is $\sim4$ and $\sim2$\,h, respectively.
The measured values of $m$ and $t_{scint}$ are listed in Table~\ref{tab:scint_pred}.

To explore the likelihood of the observed variability being caused by scintillation we compared the measured timescale and modulation index at each frequency to predictions based on relations presented by
\citet{goodman97,walker98} \citep[which are summarised in table 1 of][]{granot14} in conjunction with a model of the Galactic free electron distribution evaluated at the GRB position. 
\citet{hancock19pp} have developed an all-sky model for refractive interstellar scintillation (RISS19), which uses $H_\alpha$ maps to evaluate $SM$, $\nu_0$ and $\theta_F$ at a given observing frequency for a particularly sky position.
As the $H_\alpha$ maps have a much broader sky coverage than NE2001, which is dependent on known pulsars that cluster in the Galactic Plane, RISS19 provides better information about the $SM$ at higher Galactic latitudes. 
Note that the location of GRB 210702A is in a part of the sky  for which there is minimal pulsar coverage in NE2001 \citep[see figure 4 of ][]{cordes02}, thus our choice to use RISS19 instead.

At the position of GRB 210702A, RISS19 returns $SM=1.06 \pm 0.35 \times 10^{-3}$\,kpc\,m$^{-20/3}$ and a transition frequency of $\nu_{0}=7.66$\,GHz.
This places 9.0\,GHz near the transition frequency in the weak regime, 16.7 and 21.2\,GHz in the weak scattering regime, with 5.5\,GHz in the strong scattering regime (either refractive or diffractive). 
RISS19 also assumes that the intervening medium can be approximated as a thin screen at distance $D$, which is located half way between Earth and the edge of the Galaxy. This has been very simply modelled as a flat disk of radius 16\,kpc and height of 1\,kpc to represent the Milky Way with Earth located 8\,kpc for the Galactic centre. For the line-of-sight to GRB 210702A, RISS19 estimates a scattering screen distance of $D=0.66$\,kpc and outputs $\theta_F$ at each frequency.
Following the ISS relations summarised in table 1 of \citet{granot14}, the resulting RISS19 predicted scattering disk size ($\theta_{scatt}$), modulation index ($m_{p}$) and scintillation timescale ($t_{scint,p}$) for each frequency for the relevant scattering regime are listed in Table~\ref{tab:scint_pred}.
These include predictions at 5.5\,GHz for the refractive and diffractive scattering regimes even though we were unable to measure the modulation index or scintillation timescale from our data at this frequency. 

For the sake of comparison, the python wrapper of the {\sc FORTRAN} implementation of the NE2001 model ({\sc pyne2001}\footnote{https://pypi.org/project/pyne2001/}) returns $SM=5.5 \times 10^{-4}$\,kpc\,m$^{-20/3}$ and a transition frequency of $\nu_0=14.4$\,GHz for an extragalactic source beyond 30\,kpc. Given we observe the strongest variability at 9.0\,GHz from GRB 210702A, which is closer to the RISS19 transition frequency of 7.66\,GHz, our results seem to favour the RISS19 Galactic free electron model over NE2001.

The power of the ATCA observations of GRB 210702A is the long integration time, allowing us to probe the full scintillation cycle between 2 to 5 hours over a broad range of frequencies.  
From a quantitative perspective, the strongest modulation observed at 9.0\,GHz supports it being close to the transition frequency calculated using RISS19, with the modulation and timescale decreasing with increasing frequency according to the weak scattering relations. Given that the observed modulation index and timescales at 9.0, 16.7 and 21.2\,GHz agree with the RISS19 derived predictions to within a factor of $\sim2$, scintillation is the most likely explanation for the observed radio variability at early times from GRB 210702A. 

\subsection{Source size estimates}\label{sec:size}

\begin{table*}
    \centering
    \begin{tabular}{cccclcccc}
         \hline
         $\nu$ &  \multicolumn{2}{c}{Measured} &  \multicolumn{5}{c}{RISS19 Predictions}  \\
         \cline{2-3} \cline{5-9} 
          & $m$ & $t_{scint}$ & & Regime & $\theta_{scatt}$ & $m_p$ & $t_{scint,p}$ & Size  \\
         (GHz) &  & (hr) & & & ($\mu$as) & &(hr) & ($10^{16}$\,cm)  \\
         \hline
         5.5  & --  & --   & & SR & 7.5 & 0.8 & 4.1 & $<20$ \\
         5.5  & --  & --   & & SD$^{a}$ & 2.4 & 1 & 1.3 & $<6$  \\
         9.0  & 0.4 & 5  & & W  & 3.3 & 0.8  & 1.8 & $<9$ \\
         16.7 & 0.3 & 4  & & W  & 2.4  & 0.3 & 1.4 & $<6$ \\
         21.2 & 0.2 & 2  & & W  & 2.2  & 0.2 & 1.2 & $<6$ \\
    \end{tabular}
    \caption{The modulation index ($m$) and scintillation timescale ($t_{scint}$) measured from the light curves at each observing frequency except 5.5\,GHz  (Figure~\ref{fig:atca_lc}), which are compared to scintillation predictions from RISS19.
    At the position of GRB 210702A, RISS19 derives a transition frequency of $\nu_{0}=7.66$\,GHz with a scattering screen distance of $D=0.66$\,kpc, a scattering measure of $SM=1.06 \pm 0.35 \times 10^{-3}$\,kpc\,m$^{-20/3}$, and first Frenzel zone sizes ($\theta_{F}$) at each frequency. 
    Using these parameters, we list the size of the scattering disk $\theta_{scatt}$, 
    and the predicted modulation index ($m_p$) and scattering timescale ($t_{scint,p}$) assuming GRB 210702A is a point source at 5.5, 9.0, 16.7 and 21.2\,GHz, following scintillation relations summarised by \citet{granot14} depending on the regime (W = weak, SR = strong refractive and SD = strong diffractive). 
    An estimate of the size of the blast wave in the plane of the sky is calculated assuming an angular source size of $\theta_{scatt}$ for each frequency at the angular size distance $D_A$ derived from the redshift 1.160. \\
    $^{a}$At 5.5\,GHz, diffractive scintillation is active over a 1.8\,GHz bandwidth using equation 15 from \citet{walker98}. 
    }
    \label{tab:scint_pred}
\end{table*}

If we assume that the source of variability is dominated by ISS, it means the radio afterglow of GRB 210702A could not have had an angular size much larger than the size of the scattering disk ($\theta_{scatt}$ in Table~\ref{tab:scint_pred}).
The measured redshift to GRB 210702A of $z=1.160$ \citep{xu21} corresponds to 
an angular size distance of $D_{A}=1750$\,Mpc 
\citep[$H_0=67.4, \Omega_m=0.315$;][]{planck20}. 
The physical size of the scattering disk (diameter) is then $\theta_{scatt} \times D_{A}$, 
which corresponds to the maximum size of the jet front image on the sky. 
We consider this a proxy for the size of the blast wave in the plane of the sky  \citep{granot05}, 
listing the resulting values at each frequency in Table~\ref{tab:scint_pred}. 
At $\sim13$\,h ($\sim0.56$\,d) post-burst, we estimate the size of the blast wave on the plane of the sky was $\lesssim 6 \times 10^{16}$\,cm (note that for such an estimate we expect uncertainties of $\sim50$\%).

This ISS size upper-limit derived in the weak regime is consistent with those similarly derived for other GRBs after recalculating the jet front image size based on an angular size distance derived from the event redshift using a consistent cosmology, which all lie between $\sim1\times 10^{16} \mathrm{~and~} \sim8\times 10^{16}$\,cm \citep[GRBs 970508, 070125, 130427A, 161219B, 201216C;][]{frail97,frail00scint,waxman98,chandra08,vanderhorst14,alexander19,rhodes22}. 
However, all ISS size measurements for the above listed events were derived from radio observations observed days to weeks following the explosion, with the exception of GRB 161219B at 1.5\,d post-burst \citep{alexander19}. 
Our size measurement of GRB 210702A at $13$\,h (0.56\,d) post-burst is therefore the earliest ISS size constraint on any GRB to date, which was only possible due to the rapid multi-frequency radio follow-up over a long 11\,h integration. 

This scintillation size limit of the afterglow can be compared to model predictions of the forward shock. At such early times we are observing the afterglow emission before any jet break so we can assume the outflow is described by the \citet{blandford-mckee76} solution (closure relation discussions in Section~\ref{sec:afterglow} support this assumption). Assuming a viewing angle along the jet axis, we can calculate the radius of the GRB afterglow `image' using equation 5 from \citet{granot05}:
\begin{equation}\label{eq:radius}
    R_{\perp} = 3.91 \times 10^{16} (E_{iso,52}/n_0)^{1/8}[t_{days}/(1+z)]^{5/8} \mathrm{cm}
\end{equation}
for a homogeneous medium (assumption based on our analysis in Section~\ref{sec:afterglow}), where $E_{iso,52} = E_{iso} \times 10^{52}$\,erg is the isotropic energy of the blast wave and $n_0$ is the CBM density (cm$^{-3}$). 

To properly estimate $E_{iso}$ and $n_{0}$, one needs to perform full broadband modelling. 
However, given the rather weak dependence of the source size on these parameters, we can assume an average CBM density of $n_0 = 1$\,cm$^{-3}$, and estimate $E_{iso}$ as follows. 
There is a well defined relationship between the gamma-ray isotropic energy $E_{iso,\gamma}$ and $E_{iso}$ \citep[][]{nava14} with an efficiency factor of $\epsilon_{\gamma} \sim 0.14$ \citep[][]{beniamini15} such that 
\begin{equation}\label{eq:eiso}
    \epsilon_{\gamma} = E_{iso,\gamma} / (E_{iso,\gamma} + E_{iso}).
\end{equation}
Using $E_{iso,\gamma} \sim9.3 \times 10^{53}$\,erg derived by \citet{frederiks21}, Equation~\ref{eq:eiso} gives $E_{iso}\sim5.7 \times 10^{54}$\,erg for GRB 210702A. Substituting this value into Equation~\ref{eq:radius}, and assuming a CBM density of $n_0 = 1$\,cm$^{-3}$, the predicted forward shock radius of the GRB 210702A is $4 \times 10^{16}$\,cm (diameter of $8 \times 10^{16}$\,cm) at 0.56\,d post-burst (the time of the 9.0\,GHz light curve peak). 
If we assume that our value for $E_{iso}$ is a reasonable estimate then changing $n_{0}$ by an order of magnitude only changes $R_{\perp}$ by a factor $\lesssim1/3$. 
Changing either or both $E_{iso}$ and $n_{0}$ by an order of magnitude changes $R_{\perp}$ by a factor of $\lesssim2/3$. 
As the whole ejecta layer has a thickness of $R/\Gamma^{2}$, where $R$ is the blast wave radius, and we are still in the relativisitic regime such that $\Delta R << R$, we can assume that the reverse shock radius is comparable to the forward shock radius \citep{meszaros97,wijers99}.

This predicted forward shock radius for GRB 210702A is therefore consistent with our upper-limit of $\lesssim 6 \times 10^{16}$\,cm on the blast wave image size we derived from ISS. 
While this may indicate that our assumptions regarding the gamma-ray efficiency ($\epsilon_{\gamma} \sim 0.14$) and therefore our calculation of $E_{iso}$ are reasonable (see Equation~\ref{eq:eiso}), this relation was derived by assuming a sample of GRBs had the same micro-physical values for the fraction of thermal energy in the electrons ($\epsilon_e=0.1$) and magnetic fields ($\epsilon_B=10^{-2}$), and that $p=2.5$ \citep{beniamini15}. We already know that $p=2.9 \pm 0.1$ is likely for GRB 210702A, and while $\epsilon_e=0.1$ is reasonably well favoured to not vary much \citep[e.g.,][]{beniamini17,duncan23}, $\epsilon_B$ and the CBM density surrounding long GRBs are known to vary over several orders of magnitude \citep[e.g.,][]{cenko11,granot14}. 
It is therefore only through modelling high cadence, multi-wavelength datasets from a few hours to many months post-burst, that we can break the degeneracy between these parameters and properly test the fireball scenario for GRB afterglows using source size measurements derived from ISS. 

\section{Conclusions}

The rapid radio follow-up observations of GRB 210702A with ATCA at 5.5, 9.0, 16.7 and 21.2\,GHz detected at least one radio flare between 11 and 16\,h post-burst. The most prominent feature was detected at 9.0\,GHz, which is inconsistent with the reverse and forward shock afterglow characteristic timescales and is more likely the result of interstellar scintillation. Our analysis of the ATCA data as well as the \swift-XRT and -UVOT observations leads us to the following conclusions regarding the GRB afterglow and source of the radio variability:
\begin{enumerate}
    \item A broken power law fit to the 9.0\,GHz light curve binned on 15\,min timescales shows an extremely rapid radio brightening with temporal indices of 
    $\alpha_{1}=+7.3^{+1.7}_{-1.3}$, which quickly swaps to $\alpha_{2}=-8.4 \pm 3.7$ at a break time of $13.4 \pm 0.2$\,h ($0.56 \pm 0.01$\,d) post-burst (see Figure~\ref{fig:bpl_spec_index}). These temporal indices are $>2\sigma$ away from even the most extreme forward and reverse shock light curve predictions \citep{gao13} demonstrating the radio variability is likely caused by an extrinsic source rather than being intrinsic afterglow variability ($S_{\nu} \propto t^{\alpha} \nu^{\beta}$).
    \item Using the broken power law fit at 9.0\,GHz to interpolate the flux density values to the time of the 16.7 and 21.2\,GHz observations also demonstrated extreme spectral variability. The spectral index varied between $\beta=+3$ and $\beta=0$ between $\sim10-14$\,h ($\sim0.4$ to 0.6\,d) post-burst only to rise back to $\beta=+2$ by $\sim17$\,h ($\sim0.7$\,d) post-burst during the observed 5\,h period of flaring (see Figure~\ref{fig:bpl_spec_index}).
    \item The X-ray and optical bands are dominated by the forward shock component and are in the same afterglow power law regime below the cooling break such that $\nu_{a,m}<\nu_{X,O}<\nu_{c}$. Using the closure relations we find the CBM is likely homogeneous ($s=0$) and that $p=2.9 \pm 0.1$. Modelling the optical to X-ray SED with a single power-law spectrum and accounting for extinction effects, we derive a spectral index of $\beta_{X,O} = -0.96 \pm 0.06$, and a host galaxy extinction of $E_{B-V} = 0.069 \pm 0.009$ (Figure~\ref{fig:end_walkers}).
    \item Investigating interstellar scintillation as the source of the rapid variability, we used RISS19 \citep{hancock19pp} to calculate a transition frequency of $\nu_0=7.66$\,GHz at the Galactic coordinates of GRB 210702A. This places 9.0\,GHz just in the weak scintillation regime near the transition frequency where we would expect the largest modulation from ISS. The measured modulation and timescales of the 9.0, 16.7 and 21.2\,GHz light curves are consistent with weak scintillation predictions \citep{goodman97,walker98} within a factor of $\sim2$ (see Table~\ref{tab:scint_pred}), making ISS the most likely explanation for the observed short timescale radio variability. 
    \item Assuming the source of radio variability is dominated by ISS, the radio afterglow is unlikely to have an angular size much bigger than the scattering disk, allowing us to place an upper limit on the size of the blast wave in the plane of the sky of $\lesssim 6 \times 10^{16}$\,cm at $\sim0.56$\,d post-burst. This is consistent with forward shock source size predictions for this GRB (diameter of $8 \times 10^{16}$\,cm) and also upper limits on the source size derived using scintillation for other GRBs (between $\sim1\times 10^{16} \mathrm{~and~} \sim8\times 10^{16}$\,cm).
    \item At 0.56\,d post-burst, this is the earliest source size limit placed on a GRB blast wave to date. 
\end{enumerate}

By obtaining an 11\,h observation with ATCA swapping between dual receivers, we were able to clearly track the temporal and spectral structure of scintillation simultaneously between 4 and 22\,GHz less than 1\,d post-burst.
The clear flare structure of the light curve at 9.0\,GHz could indicate that we have observed a full scintillation cycle at this frequency.
These results argue the importance for not just rapid (within a day) radio follow-up but the need for long integrations to properly track the scintillation in time over a broad frequency range.

These results also demonstrate the power of scintillation as a direct probe of the size of the blast wave, which can then be used to test afterglow modelling assumptions and the resulting micro-physical parameters (see discussion in Section~\ref{sec:size}).  
It is also possible that the radio afterglow of GRB 210702A was only detected because of ISS, which boosted the signal above the ATCA sensitivity limit on short ($\sim15$\,min) timescales. Using these observations alone, it is not possible to tell whether we are seeing the boosting of the radio reverse or forward shock (or both) emission components. However, future ATCA rapid-response triggers on GRBs that are combined with high cadence, multi-frequency radio follow-up from a day to hundreds of days post-burst, combined with broadband modelling will allow us to disentangle the forward and reverse shock components, identify the source of the boosted emission, and allow us to learn more about the dynamics of the radio afterglow within a day post-burst.

In this paper, we have demonstrated the power of rapid-response observations for detecting short-timescale radio variability very early in the evolution of explosive transients for constraining assumptions associated with relativistic blast wave physics. This strongly motivates using the same technique for probing early-time radio variability from many other classes of transients in order to test associated emission and outflow models, and thus further supports the need for rapid-response transient triggering systems on the upcoming Square Kilometre Array.

\section*{Acknowledgements}

We acknowledge the Gomeroi people as the traditional owners of the Australia Telescope Compact Array (ATCA) observatory site. 
The ATCA is part of the Australia Telescope National Facility, which is funded by the Australian Government for operation as a National Facility managed by CSIRO. 
We acknowledge the Whadjuk Nyungar people as the traditional owners of the land on which the Bentley Curtin University campus is located, where the majority of this research was conducted.  

We thank the referee for their recommendations on this manuscript.
GEA is the recipient of an Australian Research Council Discovery Early Career Researcher Award (project number DE180100346) funded by the Australian Government.
TDR acknowledges the financial contribution from the Italian Space Agency and the Italian National Institute for Astrophysics (ASI-INAF) n.2017-14-H.0. 
AJG acknowledges support through the
Australian Research Council’s Discovery Projects funding scheme (project number DP200102471).

This work made use of data supplied by the UK {\it Swift} Science Data Centre at the University of Leicester and the {\it Swift} satellite. {\it Swift}, launched in November 2004, is a NASA mission in partnership with the Italian Space Agency and the UK Space Agency. {\it Swift} is managed by NASA Goddard. Penn State University controls science and flight operations from the Mission Operations Center in University Park, Pennsylvania. Los Alamos National Laboratory provides gamma-ray imaging analysis.

This research makes use of {\sc PyMC3} \citep{pymc3}, {\sc corner.py} \citep{corner}, {\sc Astropy}, a community-developed core Python package for Astronomy \citep{TheAstropyCollaboration2013,TheAstropyCollaboration2018}, {\sc numpy} \citep{vanderWalt_numpy_2011} and {\sc scipy} \citep{scipy} python modules. This research also makes use of {\sc matplotlib} \citep{hunter07}. 
This research has made use of NASA's Astrophysics Data System. 
This research has made use of SAOImage DS9, developed by the Smithsonian Astrophysical Observatory.

\section*{Data Availability}

The raw ATCA data are available via the Australia Telescope Online Archive\footnote{https://atoa.atnf.csiro.au/} under the project code C3374. Following standard data processing and calibration, the high-time resolution processing using \textsc{uvmultifit} was conducted using \textsc{casa} scripts that can be found at https://github.com/tetarenk/AstroCompute\_Scripts.
All results output by this analysis are recorded in Table~\ref{tab:radio_flux}.
All \swift-XRT and \swift-UVOT data used in this paper are available via the UK Swift Science Data Centre at the University of Leicester at https://www.swift.ac.uk/index.php under GRB 210702A.

\bibliographystyle{mnras}
\bibliography{papers} % if your bibtex file is called example.bib

\appendix

\section{Radio flux density measurements}

Table~\ref{tab:radio_flux} lists the flux density measurements of GRB 210702A at 5.5, 9.0, 16.7 and 21.2\,GHz as measured by \textsc{uvmultifit} in the visibility plane with integration times of 60, 15, 12.5, and 12.5 minutes, respectively. The time post-burst corresponds to the mid-time between the start and end of the combined scans for each flux density measurement, which may include overheads associated with calibrator scans and target scans at other frequencies. The bolded flux density measurements have a $SNR>3$ where the source position was not fixed during the fitting with \textsc{uvmultifit} whereas the non-bolded measurements have a $SNR<3$ and were force-fitted to the best known \swift-UVOT GRB position.

\begin{table}
    \centering
    \begin{tabular}{lccc}
         \hline
         $\nu$ & Time post-burst & Flux density & $3\sigma$ threshold \\
         (GHz) & (hr) & ($\mu$Jy/beam) & ($\mu$Jy/beam)  \\
         \hline
         5.5 
         & 6.4 & $-44 \pm 43$ & 129 \\
         & 9.1 & $60 \pm 29$ & 87 \\
         & 11.9 & $\bf{74 \pm 24}$ & 71 \\
         & 14.9 & $69 \pm 27$ & 80 \\
         \hline
         9.0
         & 5.5 & $4 \pm 82$ & 245 \\
         & 5.8 & $14 \pm 73$ & 220 \\
         & 7.0 & $-81 \pm 61$ & 183 \\
         & 7.3 & $-81 \pm 58$ & 173 \\
         & 8.5 & $28 \pm 55$ & 164 \\
         & 8.7 & $-60 \pm 110$ & 331 \\
         & 9.5 & $27 \pm 45$ & 135 \\
         & 9.8 & $-24 \pm 42$ & 126 \\
         & 10.9 & $135 \pm 43$ & 128 \\
         & 11.3 & $\bf{167 \pm 43}$ & 130 \\
         & 12.6 & $\bf{316 \pm 44}$ & 131 \\
         & 12.9 & $\bf{306 \pm 44}$ & 133 \\
         & 14.0 & $\bf{381 \pm 45}$ & 134 \\
         & 14.3 & $\bf{218 \pm 45}$ & 135 \\
         & 15.5 & $128 \pm 49$ & 146 \\
         & 15.8 & $166 \pm 50$ & 150 \\
         \hline
         16.7
         & 6.3 & $87 \pm 169$ & 506 \\
         & 6.6 & $56 \pm 164$ & 492 \\
         & 7.8 & $5 \pm 138$ & 415 \\
         & 8.1 & $8 \pm 134$ & 401 \\
         & 10.2 & $\bf{427 \pm 70}$ & 210 \\
         & 10.5 & $\bf{483 \pm 70}$ & 210 \\
         & 11.8 & $\bf{468 \pm 69}$ & 206 \\
         & 12.1 & $\bf{448 \pm 68}$ & 204 \\
         & 13.3 & $19 \pm 67$ & 200 \\
         & 13.6 & $\bf{228 \pm 67}$ & 201 \\
         & 14.8 & $\bf{257 \pm 71}$ & 213 \\
         & 15.1 & $\bf{335 \pm 73}$ & 219 \\
         & 16.3 & $\bf{300 \pm 83}$ & 249 \\
         \hline
         21.2 
         & 6.3 & $96 \pm 168$ & 504 \\
         & 6.6 & $-107 \pm 163$ & 488 \\
         & 7.8 & $49 \pm 137$ & 411 \\
         & 8.1 & $-45 \pm 133$ & 400 \\
         & 10.2 & $\bf{470 \pm 70}$ & 210 \\
         & 10.5 & $\bf{556 \pm 70}$ & 210 \\
         & 11.8 & $-59 \pm 68$ & 203 \\
         & 12.1 & $\bf{273 \pm 68}$ & 204 \\
         & 13.3 & $\bf{414 \pm 67}$ & 200 \\
         & 13.6 & $\bf{343 \pm 67}$ & 201 \\
         & 14.8 & $\bf{285 \pm 71}$ & 213 \\
         & 15.1 & $\bf{416 \pm 73}$ & 219 \\
         & 16.3 & $\bf{387 \pm 83}$ & 249 \\ 
         \hline
    \end{tabular} 
    \caption{The flux density measurements of GRB 210702A plotted in Figure~\ref{fig:atca_lc} that were obtained by
    fitting for a point source in the visibility plane using \textsc{uvmultifit}. The bolded flux densities indicate detections ($SNR>3$) and those not bolded are force-fitted flux density values ($SNR<3$). The $3\sigma$ threshold is the detection threshold and corresponds to 3 times the \textsc{uvmultifit} error. The reported time post-burst corresponds to the midpoint of the (combined) scans.}
    \label{tab:radio_flux}
\end{table}

\section{Corner plots from MCMC model fitting analyses to the 9.0\,GHz light curve and the X-ray to optical SED}

\begin{figure*}
    \centering
    \includegraphics[width=\textwidth]{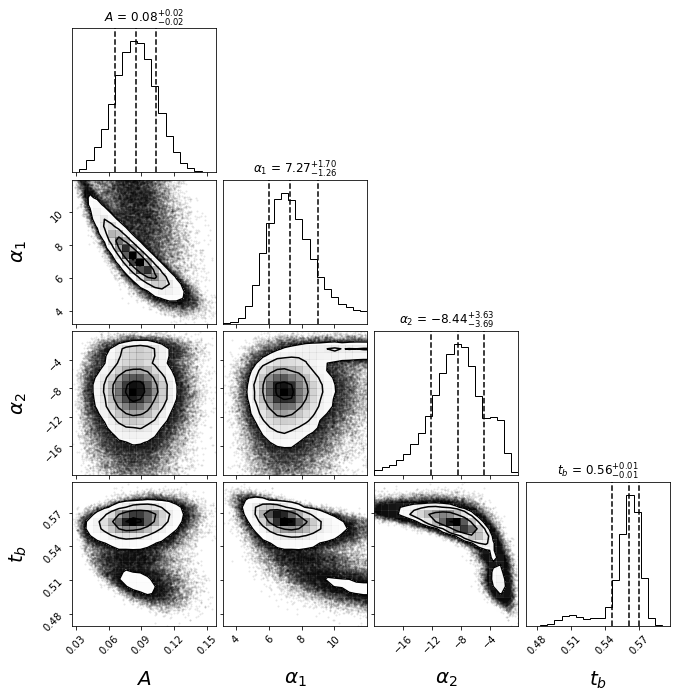}
    \caption{Corner plot for the model posterior sample from the broken power law fitted to the 9.0\,GHz light curve using the MCMC method with 10,000 step production chains for all of the 20 walkers assuming uniform priors for all four parameters. The plotted parameter $A$ is the amplitude, $\alpha_1$ is spectral index 1, $\alpha_2$ is spectral index 2, and $t_b$ is the break time (see Equation~\ref{eq:bkpl} and Section~\ref{sec:var}). The three solid contour lines on the joint distribution panels correspond to 1, 2, and 3$\sigma$ highest density intervals. The dashed lines on the marginal distribution panels show the 0.16, 0.5, and 0.84 quantiles (median and borders of a 1$\sigma$ credible interval) for each parameter.}
    \label{fig:corn_9ghz}
\end{figure*}

Figure~\ref{fig:corn_9ghz} shows the corner plots of the broken power law fit to the 9.0\,GHz light curve performed using the MCMC method as described in Section~\ref{sec:var}. As demonstrated by these corner plots and the traces in the top panel of  Figure~\ref{fig:bpl_spec_index}, there is a local minimum suggesting the possibility of an earlier power law break time. 

\begin{figure*}
    \centering
    \includegraphics[width=\textwidth]{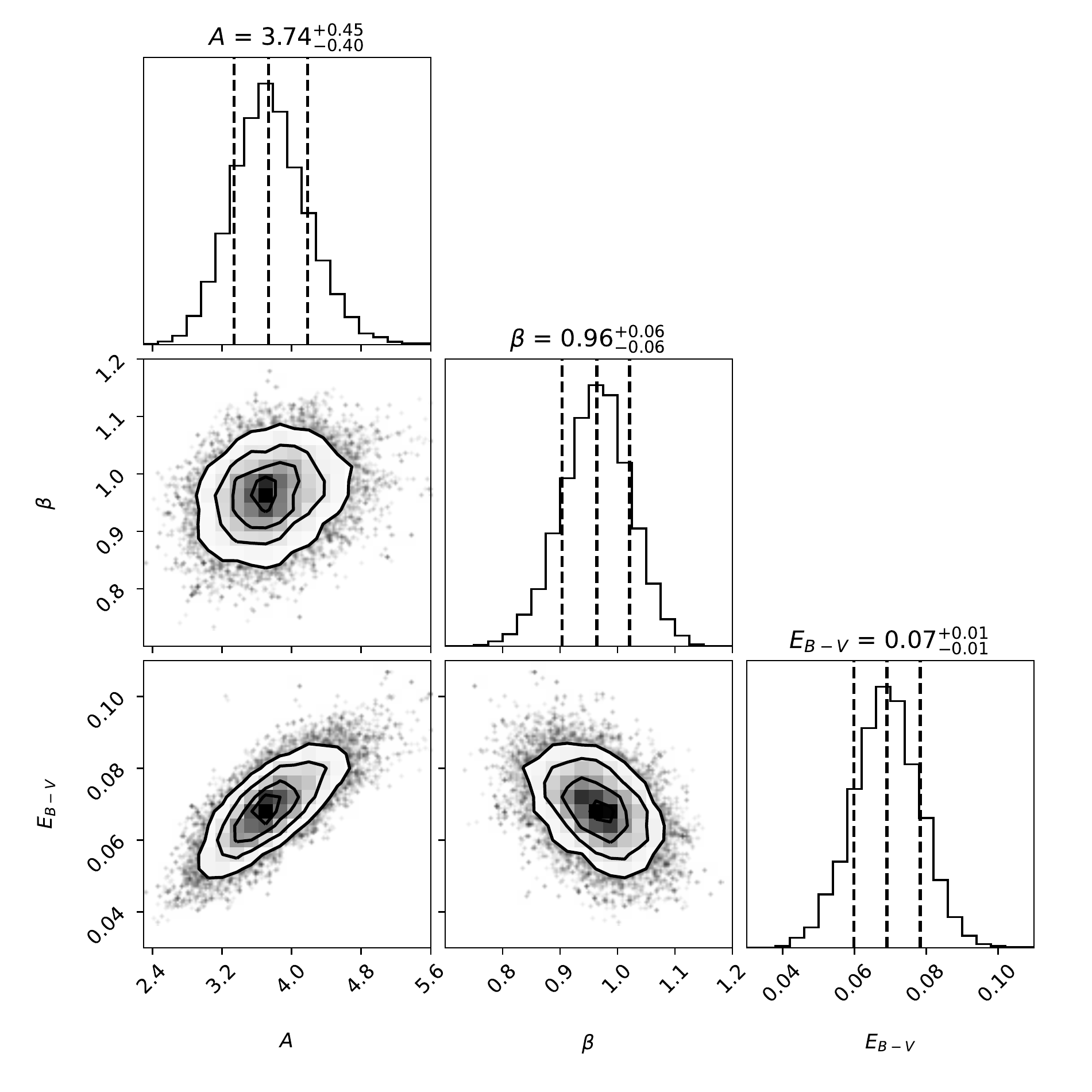}
    \caption{Corner plot from the model fitted to the X-ray to optical SED using the MCMC method with 500 step production chains for all of the 50 walkers assuming a Gaussian spectral index prior. The plotted parameter $A$ is the Milky Way extinction, $\beta$ is the X-ray to optical spectral index and $E_{B-V}$ is the host galaxy extinction (see Section~\ref{sec:swift_spec}).
    The three solid contour lines on the joint distribution panels correspond to 1, 2, and 3$\sigma$ highest density intervals. The dashed lines on the marginal distribution panels show the 0.16, 0.5, and 0.84 quantiles (median and borders of a 1$\sigma$ credible interval) for each parameter.}
    \label{fig:corn_xosed}
\end{figure*}

Figure~\ref{fig:corn_xosed} shows the corner plots of the model fitted to the X-ray to optical SED performed using the MCMC method as described in Section~\ref{sec:swift_spec}. Note that there appears to be a correlation between the amplitude and extinction, as well as an inverse correlation between spectral index and extinction, however, the extinction is well constrained.

\end{document}